\documentclass{elsart}
\usepackage{graphicx,amssymb}
\journal{Physics Letters B}

\begin{document}

\begin{frontmatter}

\title{Nuclear liquid-gas phase transition and supernovae evolution}

\author[jeje]{J\'er\^ome Margueron\corauthref{cor}\thanksref{now}}
\corauth[cor]{Corresponding author.}
\thanks[now]{Present address: Institut f\"{u}r Theoretische Physik, Universit\"{a}t
T\"{u}bingen, D-72076 T\"{u}bingen, Germany}
\ead{margueron@tphys.physik.uni-tuebingen.de}
\author[jesus]{Jes\'us Navarro}
\author[pat]{Patrick Blottiau}
\address[jeje]{GANIL CEA/DSM - CNRS/IN2P3 BP 5027 F-14076 Caen Cedex 5, France}
\address[jesus]{IFIC (CSIC - Universidad de Valencia) Apdo. 22085, E-46.071-Valencia, Spain}
\address[pat]{CEA/DIF DPTA BP12 F-91680 Bruy\`eres-le-Ch\^{a}tel Cedex, France}

\begin{abstract}
It is shown that the large density fluctuations appearing at the onset of the 
first order nuclear liquid-gas phase transition can play an important role in the
supernovae evolution. Due to these fluctuations, the neutrino gas may be
trapped inside a thin layer of matter near the proto-neutron star surface. 
The resulting increase of pressure 
may induce strong particle ejection a few hundred milliseconds after the 
bounce of the collapse, contributing to the revival of the shock wave. 
The Hartree-Fock+RPA scheme, with a finite-range nucleon-nucleon 
effective interaction, is employed to estimate the effects of the 
neutrino trapping due to the strong density fluctuations, and to 
discuss qualitatively the consequences of the suggested new scenario.
\end{abstract}

\end{frontmatter}

The first simulations of Colgate and White~\cite{Col66} and Arnett~\cite{Arn66} 
have settled the general scenario of explosive supernovae, which schematically
goes as follows. Stars with more than about ten times the Sun mass develop 
central iron cores which eventually become unstable and collapse to neutron 
stars, due to electron capture and photodisintegration onto iron-group nuclei. 
The interior of neutron stars is denser than nuclear matter and initially
extremely hot. Above nuclear density, the equation of state becomes stiff 
enough to produce a bounce of the core, and a shock wave is formed, moving to 
the infalling outer core and enveloppe. 
Particle reactions at such conditions 
create a huge number of neutrinos, which eventually escape from the dense 
neutron star, transferring energy to the matter in the still infalling outer 
layers of the star. This 
neutrino heating is believed to cause the violent disruption of the star in a 
supernova explosion. However, from the very beginning it was found that the 
resulting shock wave is not energetic enough. Recently, an enormous effort has 
been done to describe the neutrino production and interactions in great 
detail~\cite{Bur03}, including the effects of stellar rotation and of violent 
anisotropic plasma motions. Convective processes in the supernova core had 
been recognized to accelerate the energy transport inside the neutron star 
and to enhance the deposition of energy by neutrinos in the outer stellar 
layers, thus supporting the explosion of the star (the "delayed" mechanism). 
Along some decades, numerical simulations were performed, starting from 
the previous 1D codes~\cite{Bet85,Blo89} and the 
flux-limited neutrino transport scheme~\cite{Bru85} to reach the 
multi-dimensional calculations, showing the importance of convection and
Rayleigh-Taylor instabilities, and the most refined neutrino transport 
methods~\cite{Mel88,Ram02}.
Nevertheless, the outcome of these worldwide most elaborate supernova 
simulations is disappointing, as no explosions could be obtained. This negative 
result shatters the widely accepted view of how the explosion starts.

In this letter we point out that large density fluctuations associated to the first
order nuclear liquid-gas phase transition, can play an important role in the neutrino 
trapping. Indeed, the scattering of neutrinos is hugely increased at the onset 
of strong density fluctuations, so that neutrinos should be trapped
inside a thin layer of the proto-neutron star (PNS) for densities in the range 
between about $0.1\rho_0$ and $0.6\rho_0$, where $\rho_0$ 
is the nuclear matter density at saturation.
The resulting increase in the neutrino pressure may induce strong particle
ejection a few hundred milliseconds after the bounce of the collapse, 
contributing to the revival of the shock wave.  

The neutrino-nucleon cross section is obtained from a phase space integration 
of the nuclear response function. In the non-relativistic limit, the neutrino 
mean-free path $\lambda$ is given by the well-known expression~\cite{Iwa82}
\begin{eqnarray}
\frac{1}{\lambda({\rm k}_\nu,T)} = \frac{G_F^2}{16 \pi^2} \!\int \!\! d{\bf k}_3
\Bigg( c_V^2 (1+\chi)\mathcal{S}^{(0)}(q,T)
+ c_A^2 (3-\chi)\mathcal{S}^{(1)}(q,T) \Bigg)~,
\label{eq1}
\end{eqnarray}
where $k_{\nu}$ is the neutrino energy-momentum, $T$ is the temperature, 
$G_F$ is the Fermi constant, $c_V$ ($c_A$) the vector (axial) coupling constant,
${\bf k}_3$ the final neutrino momenta, $q = k_\nu-k_3$ the transferred 
energy-momentum, and $\chi=\cos\theta=\hat{\bf k}_\nu\cdot\hat{\bf k}_3$.  
The dynamical structure factors $\mathcal{S}^{(S)}(q,T)$ describe the response of 
nuclear matter to excitations induced by neutrinos, and they contain the 
relevant information on the medium. The vector (axial) part of the neutral 
current gives rise to density (spin-density) fluctuations, corresponding to 
the $S=0$ ($S=1$) spin channel. 
Density and spin-density responses contribute to the scattering
of neutrinos, while the isospin-flip response contribute to neutrino
absorption process, which is known to dominate the total cross section~\cite{Red98}.
In the following, we shall show that large isoscalar fluctuations can
give the dominant contribution at subnuclear densities.

Clearly, an enhancement of the response function in either channel will 
result in a quenching of the neutrino mean-free path. It has been 
shown~\cite{Vid84} that at densities higher than $\rho_0$ a ferromagnetic 
instability can appear, leading consequently to a zero neutrino mean-free 
path~\cite{Nav99}. However, the density at which such an instability
could appear is not yet a settled question. For instance, whereas all Skyrme 
effective interactions predict a ferromagnetic transition for densities less 
than $\sim 3.5 \rho_0$~\cite{Mar02}, recent microscopic 
calculations~\cite{fan01,vid02} push it to higher densities, in case it exists.

At present, a large nuclear community is currently studying the liquid-gas 
phase transition which occurs during multi-fragmentation experiments in 
heavy-ion collisions. 
A present status is reviewed in the recent article of Ph. Chomaz et 
al.~\cite{Cho03} and we recall the main features directly related
to our purpose.
Such a transition appears below the saturation density $\rho_0$ of nuclear 
matter, at around $\sim 0.6 \rho_0$, i.e. $\sim 0.1$~fm$^{-3}$. 
Heavy nuclei are located on the liquefaction curve for $T$=0 MeV 
and the critical temperature is located between 15 and 20 MeV, depending on 
the specific nuclear model used. The onset of liquid-gas coexistence phase is 
associated to the spinodal instability, related to a negative curvature in the 
nucleonic free energy density such that the speed of sound vanishes. 
The analysis of the topology of the thermodynamic potential show that
the phase transition is first order in symmetric and asymmetric nuclear matter.
A unique spinodal instability is expected and the asymmetry do not modify its
properties~\cite{Bar01,Mar03}. 
The spinodal instability is isoscalar (chemical instability), because the 
interaction between protons and neutrons is attractive at low density.
In the density regime close to the spinodal, the density response function
will be enhanced, thus affecting the un-charged current neutrino opacity
while the charge current neutrino opacity is not modified.
This may result in a reduced total neutrino mean-free path.

\begin{figure}[tb]
\center
\includegraphics[scale=0.3]{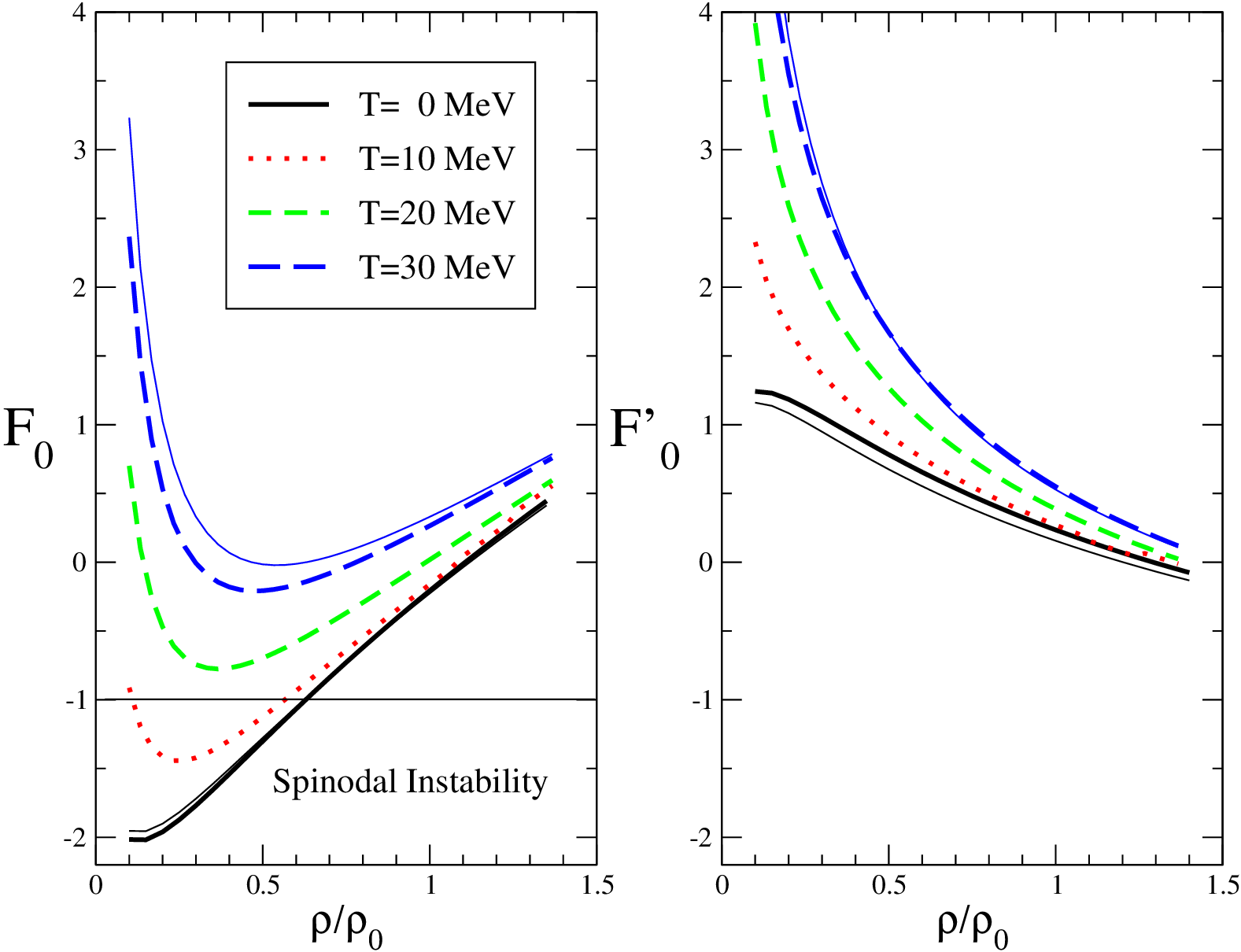}
\caption{(color online) The adimensional Landau parameters $F_0$ 
and $F^\prime_0$ are plotted 
as a function of the density at several values of temperature. 
The thin line represent a calculation of the same quantity in asymmetric
nuclear matter with $x_p=0.25$ and for $T=0$ and $30$ MeV.
The definitions 
$F_0=(F_0^{nn}+F_0^{pp}+2F_0^{np})/2$ and 
$F^\prime_0=(F_0^{nn}+F_0^{pp}-2F_0^{np})/2$ 
has been taken and the coefficients $F_0^{\tau\tau\prime}$ 
are the second derivative of the free energy density with respect to the densities. 
The onset for the spinodal instability is given by $F_0=-1$.}
\label{figa4}
\end{figure}

It is worth pointing out that Sawyer obtained a quite similar result, in his
investigation on the dependence of the neutrino opacity and the
equation of state of pure neutron matter~\cite{Saw75}. 
Assuming that the neutrino scattering is dominated
by the classical fluctuations, he concluded that a local softening of the equation
of state could produce a thin layer, near to the star surface, quite opaque
to neutrinos of some energy. In this work we consider asymmetric nuclear 
matter and we substantiate Sawyer's crude estimate
by calculating consistently both the density fluctuations and the nuclear equation
of state using the Gogny D1P finite-range nucleon-nucleon effective 
interaction~\cite{Far99}.
We describe the nuclear isoscalar and isovector density fluctuations within the
Hartree-Fock (HF) plus the Random Phase Approximation (RPA).
The calculation of the RPA response functions is
performed in the Landau approximation, ignoring the Landau coefficients
beyond the dipole one in both density and spin channels~\cite{Mar01}. 
It is worth recalling that some values of Landau parameters are related to
different instabilities. In particular, the density at which 
the dimensionless Landau parameter $F_0$ equals -1 indicates the onset of
the spinodal instability. 
In Fig.~\ref{figa4}, we show the Landau parameters $F_0$ and $F^\prime_0$ versus 
the total density for symmetric nuclear matter and several densities.
The thin line correspond to asymmetric nuclear matter with 
$x_p\equiv Z/(N+Z)=0.25$. The results in asymmetric nuclear matter look very
similar and we clearly see that out of the spinodal region, $F_0$ is attractive 
for a wide range of densities and temperatures.
The Landau parameter $F^\prime_0$ which is responsible to isospin-flip fluctuations
do not show any instable behaviour, which confirms that the instability is
isoscalar.

\begin{figure}[tb]
\center
\includegraphics[scale=0.3]{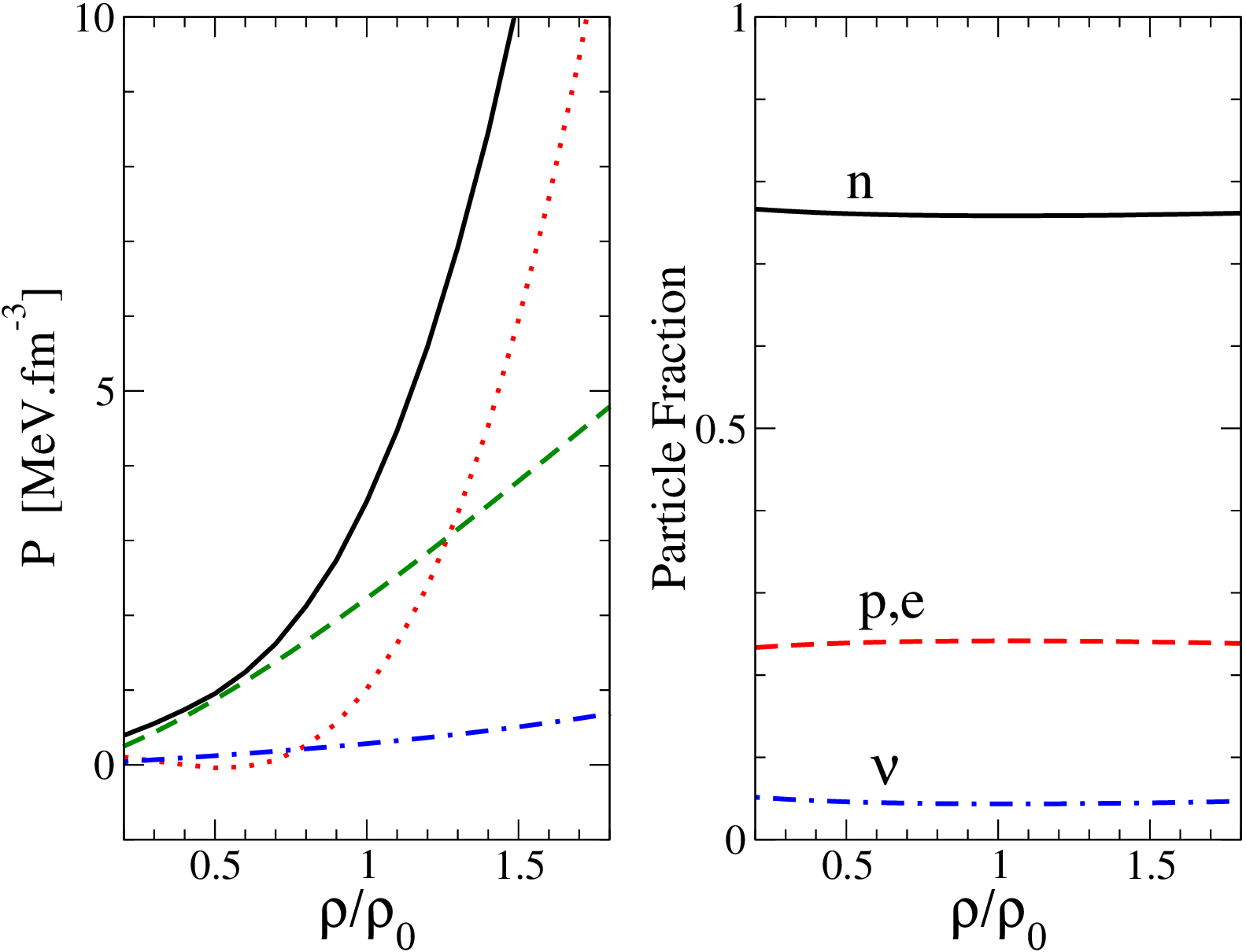}
\caption{(color online) Pressures (left panel) and particle fractions (right panel) 
plotted as a function of the density for asymmetric nuclear matter at 
temperature T=10~MeV and leptonic fraction $Y_l$=0.285, calculated with the 
Gogny D1P effective interaction.
The total pressure (solid line) is the sum of the nuclear (dotted line), the 
electronic (dashed line) and the neutrino (dot-dashed line) pressures.
On the right panel, solid, dashed and dot-dashed lines refer to the particle 
fraction of neutrons, charged particles (i.e. protons and electrons), and 
neutrinos, respectively.}
\label{figa1}
\end{figure}

In Fig.~\ref{figa1} we plot the pressures and particle fractions as a
function of the density for asymmetric nuclear matter at a temperature of
$T$=10~MeV and leptonic fraction $Y_l$=0.285, defined as 
$(\rho_e+\rho_\nu)/\rho$. 
In the low density region $\rho<\rho_0$, the total pressure is mainly
supported by electrons, and the contribution of neutrinos is about 
20\% or less~\cite{Sur84}. Hence, an increase of the neutrino pressure
by a factor of 5 will increase by a factor of 2 the total pressure.

Let us assume the extreme hypothesis that, after a few hundred milliseconds, 
all the neutrinos contained inside the star are trapped inside the thin layer 
near the surface, and that the initial distribution of neutrino density is 
quasi-uniform. Therefore, the increase of the neutrino pressure inside the thin 
layer is given by the ratio of the PNS radius $R$ over the width 
$\epsilon$ of the thin layer and writes
\begin{equation}
\frac{P_{\rm layer}}{P_{\rm inside}} \sim \left( \frac{R}{3\epsilon} 
\right)^{4/3}
\end{equation}
where the neutrino pressure is polytropic versus the density with the power $4/3$.
Hence, for a typical radius of 10~km and a width of 100~m, the neutrino 
pressure increases by a factor greater than 200. 
Of course, the Pauli principle should strongly moderate this factor, but
this schematic calculation points out the possible efficiency of the 
neutrino trapping.

This qualitative calculation assumes that all species are 
independent from each other. In fact, neutrinos are in $\beta$-equilibrium, 
so that a modification of their number implies a corresponding change of 
the number of other particles, in particular the electrons. 
If the neutrino density is increased at constant temperature and constant 
baryonic density, then 
the $\beta$ equilibrium shifts in the direction to raise the number
of electrons and protons. This is the consequence of Le Chatelier's principle of
minimum constraint extended to quantum systems. 
The same principle predicts that when the temperature is
increased, the system tends to move towards the endothermic
direction, thus increasing the number of electrons and protons.
Consequently, raising both neutrino density and temperature
produces electrons and protons.
As electrons are mainly responsible for the pressure at low density, the
overall effect of neutrino trapping is the increase of the total pressure.

\begin{figure}[tb]
\center
\includegraphics[scale=0.3]{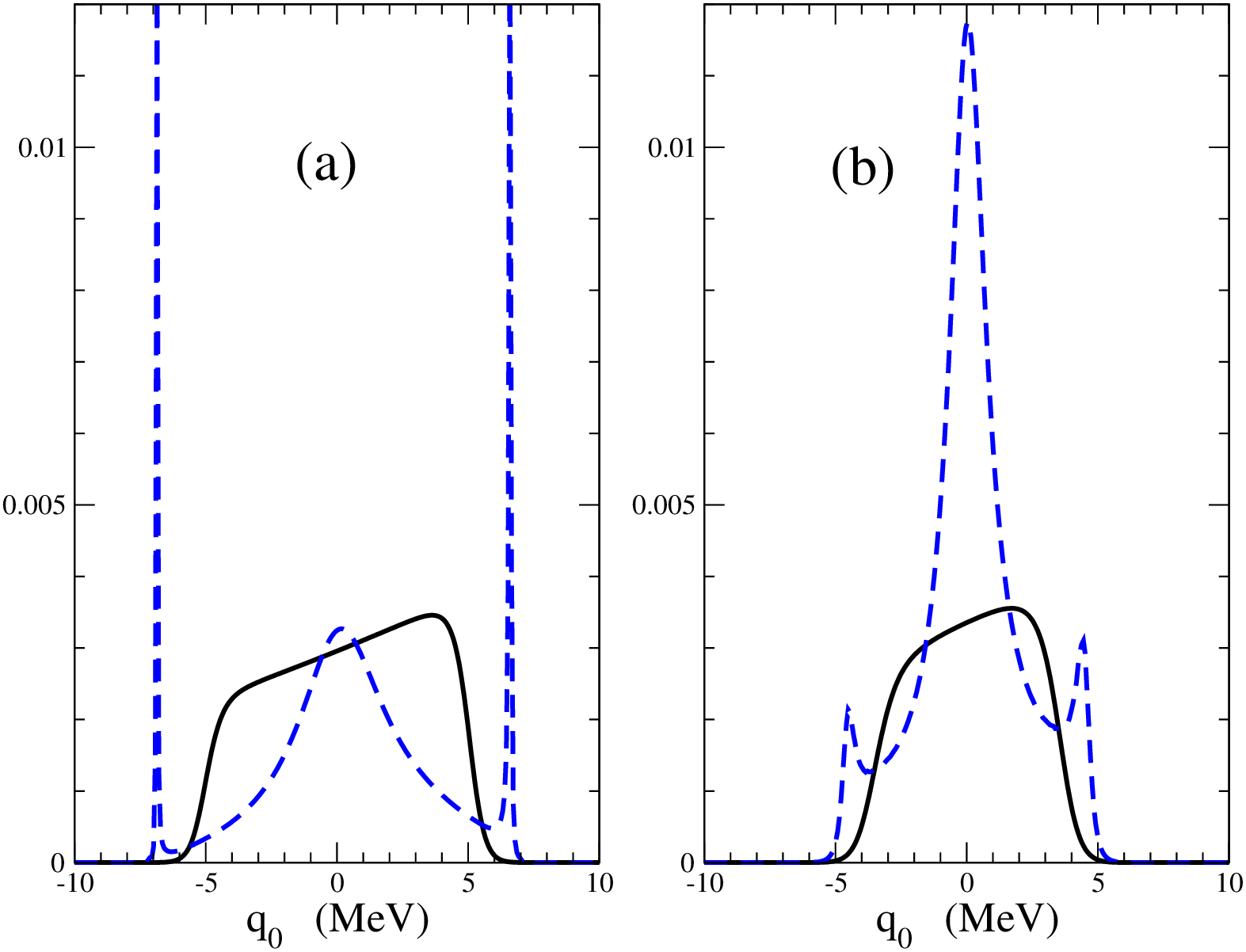}
\caption{(color online) The density-density response function 
$\mathcal{S}^{(0)}(q,T)$ is
plotted as a function of the transfered energy for two values of the
density: (a) $\rho_0$ and (b) 0.75$\rho_0$. The calculations have been done for
momentum $q=10$~MeV, temperature $T=10$~MeV and leptonic fraction $Y_l$=0.285.
The solid and dashed lines stand for the HF and the RPA results respectively.}
\label{figa2}
\end{figure}

Now we concentrate on the mechanism responsible for neutrino trapping.
In the following, we will approach the liquid-gas phase transition from the 
homogeneous nuclear matter. This can be done by decreasing either the density at
constant temperature or the temperature at constant density.
In both cases, we should keep the density beyond the spinodal one or the
temperature beyond the critical one.
The density-density response function $\mathcal{S}^{(0)}(q,T)$ is represented in
Fig.~\ref{figa2} for density values approaching the spinodal point,
namely $\rho_0$ (panel a) and 0.75$\rho_0$ (panel b). The solid line stands 
for the HF calculation while the dashed line includes the RPA correlations, 
with a particle-hole interaction described in terms of the monopolar and dipolar
Landau parameters. 
In the left panel, we see two symmetric collective states inside 
the continuum strength. These states indicate that near the Fermi surface, the
residual interaction is repulsive.
It can be seen that when the system approaches the 
liquid-gas phase transition, a zero energy mode becomes increasingly enhanced.
In fact, the peak diverges at the spinodal density. 
This is a signature of a strong attractive residual interaction, which
indeed induces the phase transition.

\begin{figure}[tb]
\center
\includegraphics[scale=0.3]{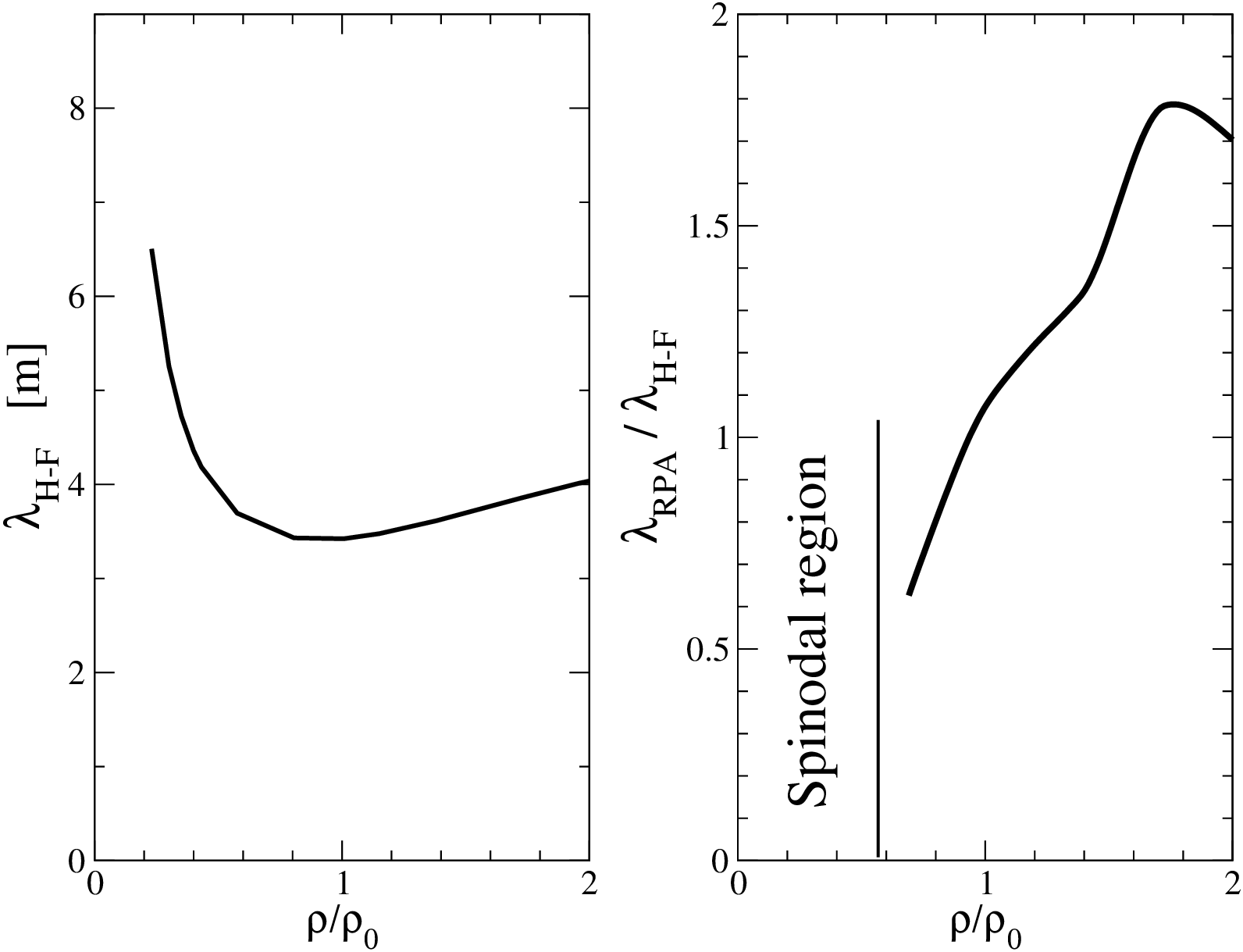}
\caption{The neutrino mean-free path is plotted as a 
function of the density for
asymmetric nuclear matter at $T$=10 MeV and $Y_l$=0.285. Left pannel: HF result,
right pannel: ratio between RPA and HF mean-free paths.}
\label{figa3}
\end{figure}

The neutrino mean-free path as a function of the density is plotted in 
Fig.~\ref{figa3} for a temperature $T$=10 MeV and a leptonic fraction
$Y_l=0.285$. Left and right panels correspond to HF and RPA results,
respectively. One can see that the RPA correlations induce a strong reduction 
of the neutrino mean-free path inside the liquid-gas phase transition in
agreement with the previous results of R. Sawyer~\cite{Saw75}. 
Hence, neutrinos could be massively trapped inside the thin external layer with 
densities near the densities of the liquid-gas phase transition, and a 
radius close to the radius of the proto-neutron star. 
The spinodal density is close to 0.6$\rho_0$ in symmetric nuclear matter and 
is nearly constant in asymmetric nuclear matter until $\rho_x/\rho\sim 0.1$ 
($x=n$ or $p$)~\cite{Mar03}.

Of course, the RPA scheme can be questionable in such a large fluctuations 
regime. The result of effects as short-range correlations, possible 
contributions of two or more loop processes at these densities have to be
elucidated, as well as the response to neutrino probes of inhomogeneous nuclear 
matter inside the phase coexistence. Nevertheless, we believe that our 
approach reflects the basic physics near the spinodal point.

Let us now analyze the physical consequences for the supernovae-II mechanism 
due to the interplay between neutrino propagation and the liquid-gas phase 
transition. During the iron core collapse, neutrino are trapped inside the core.
The energy gained by the collapse is essentially carried away by the neutrinos 
after the bounce. Neutrinos are massively trapped for densities below the 
saturation one, and hence those neutrinos which are diffusing from the inner 
part of the PNS cannot reach the outer part of the star. An important fraction 
of neutrinos
accumulates inside a thin layer close to the surface of the PNS. The neutrino 
Fermi energy increases and shifts the beta equilibrium to the production of 
protons and electrons. The neutrino and electronic pressures increase as well 
as the temperature inside the thin layer. We believe that this effect may lead 
to a partial conversion of the energy carried away by the neutrino gas to the 
matter through the beta equilibrium. According to the Euler transport equation, 
a local increase of pressure will produce an ejection of matter. This radiated 
matter will encounter the shock wave and may contribute to its revival. 
Indeed, the energy carried away by neutrinos represents about 98\% of the 
collapse energy while the shock wave transports only 2\% of it. If the ejection 
of particles carries away only 2\% of neutrino energy, the total kinetic energy 
of matter will increased by a factor of 2. This is the reason why we believe that 
this new scenario may contain an efficient mechanism for the explosion.

This scenario may modify the neutrino emission signal from supernovae.
Indeed, at the beginning of the collapse (for $\rho\sim$10$^{10}$g/cm$^3$), 
an electronic neutrino burst should come from the electronic capture.
Then neutrinos are trapped and do not radiate from the PNS. 
After the bounce, if the proposed mechanism is strong enough to produce 
a second shock wave, the neutrinos ''englued'' inside the thin layer would
be massively liberated. In this case, a second neutrino burst,
mixing the leptonic flavors, would be observed.
The observation of two neutrino bursts could be a signature of the second
explosion. 
However, this scenario do not necessary imply a second shock wave but could 
produce only a huge ejection of matter. Then, we stress that the non-observation
of two neutrino bursts is not in contradiction with this scenario.

\smallskip
In this letter we have explored the consequences that large density fluctuations
inside the liquid-gas phase transition could have for neutrino
propagation and supernovae-II explosion. 
The calculations presented here are very schematic and apply only to the homogeneous
phase close to the coexistence region, either for densities higher than the spinodal
one, or temperatures higher than the critical one.
For instance, close to the critical temperature, we expect to observe the coherent 
scattering of neutrinos in analogy with the well-known critical opalescence phenomenon 
(with respect to photons) observed in condensed matter.
Recently, neutrino opacity has also been found to become very high inside the pasta
phase, due to the coherent scattering of neutrinos off clusters~\cite{Hor04}. 
This indicate that inhomogeneous nuclear matter should also trapp neutrinos,
and increase the efficiency of our scenario.
In conclusion, this article focuses on neutrino propagation close to the liquid-gas
phase transition in nuclear matter and shows that not trivial medium effects can
trapp neutrinos. Based on this effect, we propose to include the trapping in
hydrodynamical simulations.
Work in this direction is now in progress.

{\bf Acknowledgments}

Interesting discussions with M. Baldo, Ph. Chomaz, H. M\"{u}ther,
A. P\'erez-Ca\~nellas and A. Sedrakian are gratefully acknowledged. 
J.M. wants also to thank N. Van Giai and late D. Vautherin 
for introducing him into the subject of neutrino propagation
in nuclear matter. 
J.N. is supported by MCyT/FEDER (Spain), grant number BMF2001-0262,
and Generalitat Valenciana, grant number GV01-216.

\end{document}